\documentclass[conference,9pt]{IEEEtran}
\IEEEoverridecommandlockouts
\usepackage{cite}
\usepackage{amsmath,amssymb,amsfonts}
\usepackage{algorithmic}
\usepackage{graphicx}
\usepackage{textcomp}
\usepackage{booktabs}
\usepackage[dvipsnames]{xcolor}

\usepackage{amssymb}    
\usepackage{array}    
\usepackage{pifont}     

\usepackage{lipsum}

\let\originallipsum\lipsum
\renewcommand{\lipsum}[1][]{\color{gray}\originallipsum[#1]\color{black}}
\usepackage{url}

\newcommand{\cmark}{{\color{green!60!black}\checkmark}}
\newcommand{\xmark}{{\color{red}\ding{55}}}

\usepackage{tikz}

\newcommand*\circled[1]{\tikz[baseline=(char.base)]{
            \node[shape=circle,draw,inner sep=0.8pt, minimum size=2pt] (char) {#1};}}
\newcommand{\rpoint}[1]{\circled{{\fontfamily{pcr}\selectfont\footnotesize{#1}}}}

\newcommand{\tp}[1]{\mathrm{True}(#1)}
\newcommand{\ep}[1]{\mathrm{Est}(#1)}

\usepackage{multirow}
\usepackage{subcaption}

\usepackage{fancyhdr}
\fancypagestyle{firstpage}{
  \fancyhf{}
  \fancyhead[C]{ 	To appear at the 2025 International Conference on Computer-Aided Design (ICCAD), Oct 26-30, Munich}
 
}

\begin{document}
\title{ApproxGNN: A Pretrained GNN for Parameter Prediction in Design Space Exploration for Approximate Computing
\thanks{This work has been supported by the Czech Science Foundation grant 24-10990S.}
}

\author{\IEEEauthorblockN{Ondrej Vlcek}
\IEEEauthorblockA{\textit{Brno University of Technology}\\
Brno, Czech Republic \\
xvlcek27@stud.fitbr.vut.cz\vspace*{-0.8cm}}
\and
\IEEEauthorblockN{Vojtech Mrazek}
\IEEEauthorblockA{\textit{Brno University of Technology}\\
Brno, Czech Republic \\
mrazek@fit.vutbr.cz\vspace*{-2cm}}
}


\maketitle
\thispagestyle{firstpage}

\begin{abstract}
Approximate computing offers promising energy efficiency benefits for error-tolerant applications, but discovering optimal approximations requires extensive design space exploration (DSE). Predicting the accuracy of circuits composed of approximate components without performing complete synthesis remains a challenging problem. Current machine learning approaches used to automate this task require retraining for each new circuit configuration, making them computationally expensive and time-consuming. 

This paper presents ApproxGNN, a construction methodology for a pre-trained graph neural network model predicting QoR and HW cost of approximate accelerators employing approximate adders from a library. This approach is applicable in DSE for assignment of approximate components to operations in accelerator. Our approach introduces novel component feature extraction based on learned embeddings rather than traditional error metrics, enabling improved transferability to unseen circuits. ApproxGNN models can be trained with a small number of approximate components, supports transfer to multiple prediction tasks, utilizes precomputed embeddings for efficiency, and significantly improves accuracy of the prediction of approximation error. On a set of image convolutional filters, our experimental results demonstrate that the proposed embeddings improve prediction accuracy (mean square error) by 50\% compared to conventional methods. Furthermore, the overall prediction accuracy is 30\% better than statistical machine learning approaches without fine-tuning and 54\% better with fast fine-tuning. The proposed methodology effectively addresses the challenge of transferring knowledge across different circuit designs without requiring expensive retraining. We provide our implementation, including the graph generation tools and pretrained models, as an open-source library to facilitate further research in this area.
\end{abstract}


\section{Introduction}
Approximate computing has emerged as a promising paradigm for developing highly energy-efficient computing systems and hardware accelerators for applications such as image filtering, video processing, and data mining \cite{axcomp}. This approach capitalizes on the inherent error resilience of many applications to trade Quality of Result (QoR) with energy efficiency. At the circuit level, functional approximation is achieved by employing approximate implementations for carefully selected operations within an accelerator.

For functional approximation, researchers have developed libraries of well-characterized approximate components such as adders or multipliers \cite{Hanif2017QuAd, Mrazek2017EvoApprox8b:Methods}. These libraries contain thousands of components with different trade-offs between accuracy and hardware cost. When approximating a given accelerator (in our case, a combinational circuit, as memories are not considered), designers must bind components from the library to the operators in the circuit. This problem is typically transformed into a design space exploration (DSE) challenge. Given the enormous number of possible combinations, an efficient search is typically guided by a surrogate model that can quickly predict the accuracy and hardware costs of candidate solutions~\cite{Mrazek2019AutoAx:Components, Zhang2024ApproxPilot:Framework}. 

However, predicting the accuracy and efficiency of circuits composed of approximate components without performing complete synthesis presents significant challenges. Current evaluation methods are computationally expensive and time-consuming. Existing machine learning (ML) approaches for accuracy prediction utilize principles of supervised learning -- creating training and test data sets (including feature selection, evaluation), model training, and validation. These approaches require training new models for each target accelerator, severely limiting their practical utility in real-world design flows \cite{Zhang2024ApproxPilot:Framework, Prabakaran2023Xel-FPGAs:Systems}.

Several state-of-the-art works have addressed this challenge with varying degrees of success. Design space exploration guided by ML has shown promising results \cite{Mrazek2019AutoAx:Components} in the exploration of image filters (Sobel edge detector or Gaussian filters). This approach has been extended to FPGAs and multi-stage designs of signal-processing system \cite{Prabakaran2023Xel-FPGAs:Systems}. However, these methods typically employ simple statistical ML models that must be \textbf{retrained for every new accelerator} design. More advanced approaches have been proposed in the area of high-level synthesis (HLS) for accurate circuits \cite{Alcorta2023SpecialDesign, Wu2022High-LevelAdvancing, Ustun2020AccurateNetworks}, using graph neural networks (GNNs) to predict parameters such as delay, area, or power consumption.

These GNN techniques have been successfully applied to approximate circuits as well. Zhang et al. \cite{Zhang2024ApproxPilot:Framework} employed GNNs for error prediction and outperformed statistical models defined in earlier work \cite{Mrazek2019AutoAx:Components}. However, this approach still lacks reusability -- it requires creating new training sets based on random bindings, and these datasets must be fully evaluated using synthesis and simulation tools. Moreover, these models typically rely only on predefined \textbf{error metrics of components as features} (i.e., mean arithmetic error, worst-case error, etc.), rather than learning more expressive representations.

\begin{table}[ht]
  \centering
  \caption{Comparison of the existing tools for HW accelerators' parameters prediction.}
  \label{tab:soa}
\setlength{\tabcolsep}{1pt} 

  \resizebox{\linewidth}{!}{\begin{tabular}{lccccc}
    \toprule
    \textbf{Feature} & \textbf{AutoAx} & \textbf{HW2VEC} & \textbf{HLS} & \textbf{ApproxPilot} & \textbf{ApproxGNN} \\
    & \cite{Mrazek2019AutoAx:Components} & \cite{Yu2021HW2VEC:Security} & \cite{Alcorta2023SpecialDesign, Wu2022High-LevelAdvancing, Ustun2020AccurateNetworks} & \cite{Zhang2024ApproxPilot:Framework} & \textbf{(Proposed)} \\
    \midrule
    Neural network models & \xmark & \cmark & \cmark & \cmark & \cmark \\
    HW parameters & \cmark & \xmark & \cmark & \cmark & \cmark \\
    QoR prediction & \cmark & \cmark$^*$ & \xmark & \cmark & \cmark \\
    Embeddings & \xmark & \cmark & \cmark & \xmark & \cmark \\
    Component-level repr. & \cmark & \xmark & \xmark & \cmark & \cmark \\
    Without retraining  & \xmark & \cmark & \cmark & \xmark & \cmark \\
    Design space exploration & \cmark & \xmark & \xmark & \cmark & \cmark \\
    \bottomrule
  \end{tabular}}%
  \\  \smallskip
  \raggedright
  $^*$ Security-focused parameters detection only
  \vspace*{-1em}
\end{table}

Alternative approaches have suggested that using more sophisticated \textbf{feature extraction} methods could provide significant improvements. Ansari et al. \cite{Ansari2019ImprovingMultipliers} demonstrated that the approximation error in neural networks correlates with multiple error metrics rather than a single metric. At the RTL level, specialized embeddings proposed in HW2VEC \cite{Yu2021HW2VEC:Security} help extract features for identifying security issues in EDA workflows.

Current methods have several critical limitations that hinder their practical application:

\textbf{Retraining overhead:} As mentioned in AutoAx \cite{Mrazek2019AutoAx:Components}, constructing 4,000 random solutions for training the models (running synthesis and evaluation for every solution) takes approximately 11 hours, while constructing the Pareto front takes 3 hours, and final evaluation requires additional 3 hours on high-end CPU. If we could skip the first step, we could achieve a 65\% speedup for each input hardware accelerator.

\textbf{Limited datasets:} Publicly available datasets \cite{openlsdfg,chowdhury2021openabc,Li2023InvitedDesigns} for training ML models are focused on logic synthesis but do not support approximate circuits with different QoR. Moreover, randomly generated approximate datasets tend to be biased towards less accurate solutions.

\textbf{Manual feature engineering:} Existing approaches often rely on manually defined features that are application-specific and may not generalize well across different designs.

Therefore, improving the reusability and transferability of prediction models for approximate computing is one of the major research challenges for realizing efficient design space exploration. The overall comparison of existing approaches and the propose tool is given in Tab. \ref{tab:soa}.

In this paper, we present \textbf{ApproxGNN}, a pretrained Graph Neural Network for parameter prediction in design space exploration for approximate computing. Unlike previous approaches, ApproxGNN eliminates application-specific training, offering an immediately usable prediction model that can generalize across different approximate computing scenarios. Our approach leverages learned embeddings for component feature extraction, providing richer representations than conventional error metrics.

Our novel contributions include:
(i) a \textbf{novel component feature extraction} method based on learned embeddings that captures the functional characteristics of approximate components more effectively than traditional error metrics, (ii) a \textbf{Verilog parser for GNNs} that transforms Verilog code into the desired graph representation, supporting various types of accelerators through a grammar extraction method in Python, (iii) a \textbf{new method for generating graphs} of synthetic graphics kernel circuits, enabling the creation of comprehensive training datasets, and (iv) a \textbf{pretrained universal model} for feature extraction from approximate components that is applicable to different tasks without retraining, improving transfer performance over traditional error metrics.

Our experimental results demonstrate that ApproxGNN's embeddings improve the prediction accuracy (mean square error) by 50\%, and the overall prediction accuracy is by 30\% and 54\% better than the statistical ML without and after fast fine-tuning, respectively.

An open-source library containing the Verilog parser, dataset, and resulting models is available at \textcolor{blue}{\url{https://github.com/ehw-fit/approx-gnn}}.

\section{Background and Related Work}

\subsection{Design Space Exploration for Approximate Computing}
\begin{figure}[b]
     \centering\vspace{-0.8em}
     \includegraphics[width=0.9\linewidth]{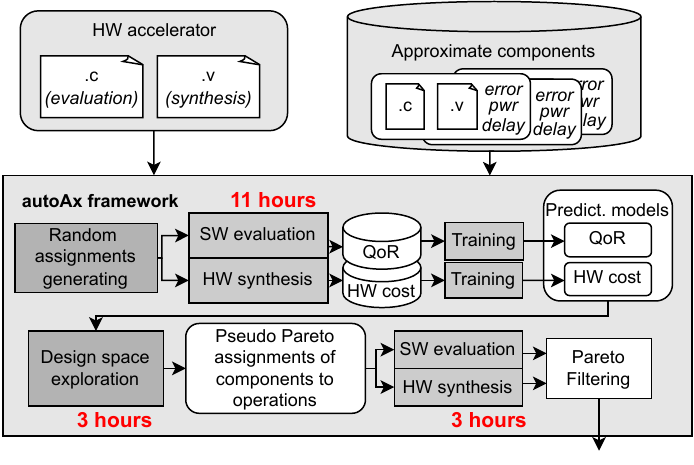}\vspace{-.8em}
     \caption{Overall flow of AutoAx~\cite{Mrazek2019AutoAx:Components} DSE algorithm guided by ML models. The red labels show the time needed for the key parts of the algorithm.}
     \label{fig:autoax}
 \end{figure}

Mrazek et al.~\cite{Mrazek2019AutoAx:Components} introduced AutoAx, a methodology for selecting circuits from libraries of approximate components to construct complex accelerators showing a good tradeoff between the QoR and HW costs. The key innovation in this approach is the use of ML techniques to create models that estimate the overall QoR and HW cost without performing full synthesis at the accelerator level. AutoAx utilized relatively simple ML models like random forest or decision trees that are better than naive approaches. A significant limitation of this approach is that model training requires pre-evaluated datasets. These datasets consists of thousands of random configurations, that are created for every input accelerator and fully evaluated using simulation and synthesis tools, resulting in very long adaptation times. These models guide the design space exploration algorithm with fast estimation of the QoR and hardware cost. Fig.~\ref{fig:autoax} shows that the construction of training data took 11 hours and needs to be re-run for every input accelerator.

Prabakaran et al.~\cite{Prabakaran2023Xel-FPGAs:Systems} extended this work in their Xel-FPGAs framework by adapting for FPGAs and adding additional features for the ML model. Instead of using parameters of components after FPGA synthesis, they employed fast ABC synthesis systems. Moreover, they proposed a multi-stage algorithm that creates more advanced approximate building blocks (such as filters) from basic arithmetic components, which are then integrated into the final application. However, this approach still faces the challenge that ML algorithms need to be trained on application-specific datasets, requiring significant computational resources and time for each new design scenario.

\subsection{Neural Networks for Hardware Parameter Prediction}
Simple statistical ML techniques are not the only way to predict circuit parameters~\cite{Alcorta2023SpecialDesign}. Recent research has demonstrated that GNN-based approaches can significantly improve prediction accuracy and generalization capabilities. GNNs are neural architectures operating on graph-structured data. These networks iteratively update the node representations by aggregating the representations of node neighbors and their own representation in the previous iteration \cite{Wu2022dnn}.

Ustun et al.~\cite{Ustun2020AccurateNetworks} proposed a GNN approach for delay prediction in High-Level Synthesis (HLS). Although not focused on approximate computing, they demonstrated a viable approach using GNNs to guide the mapping of circuits to LUTs and DSPs for delay prediction. Building on this concept, Wu et al.~\cite{Wu2022High-LevelAdvancing} employed hierarchical training for mapping hardware resources (LUT, DSP, FF) and subsequently used a second model for graph-level regression. Their approach successfully generalized to unseen real-world programs.

For ASIC circuits, Xu et al.~\cite{Xu2022SNSsPredictor} proposed SNS, a deep learning-based framework for predicting area, power, and timing. Their approach involved extracting graphs and circuit paths from Verilog designs and training models using various benchmark circuits at the RTL level, including registers. SNS achieved prediction speeds two to three orders of magnitude faster than traditional synthesis tools while maintaining reasonable accuracy. Large datasets for training the predictors of synthesis parameters (obtained by \textit{abc} synthesis tools) were published \cite{openlsdfg,chowdhury2021openabc}. However, these comprehensive datasets focus only on the RTL level of description and do not consider the QoR parameters since they target accurate circuit design.

\subsection{Error Prediction for Approximate Computing}
In the domain of approximate computing, error prediction is critical for ensuring that approximations maintain acceptable accuracy levels. Vaeztourshizi et al.~\cite{Vaeztourshizi2023EfficientSystems} proposed an error modeling approach for library modules by dividing input ranges into intervals and characterizing output errors for different combinations of these intervals. Their method relied on statistical calculations rather than ML techniques.

Mo et al.~\cite{Mo2024LearningApplications} demonstrated the advantages of ML for error prediction in approximate multipliers. They manually selected features from approximate components and employed various ML algorithms (SVM, MLP, DT, and RF) for prediction. Their approach aimed to estimate accuracy loss for neural network inference and showed better performance for regularly approximated multipliers with techniques like path cutting and cell omission.

\subsection{Graph-based Representations for Hardware}

The discontinuous nature of hardware designs presents unique challenges for applying machine learning techniques. 
Yu et al.~\cite{Yu2021HW2VEC:Security} introduced HW2VEC, an open-source graph learning tool specifically designed for hardware security applications. This framework provides an automated pipeline for extracting graph representations from hardware designs at different abstraction levels (both register-transfer level and gate-level netlists) with specialized embeddings. However, this pipeline requires all components to be flattened into a single circuit, making it unsuitable for tasks where only component parameters are known and inserted. Its primary focus on distinguishing broad HW functionality also makes the embeddings insufficient for precise regression tasks. Another approach employing the GNNs, the Verilog-to-PyG framework~\cite{Li2023InvitedDesigns}, provides an interface between EDA tools for augmentation and the PyTorch Geometric graph learning platform. 

\subsection{GNNs for Approximate Computing DSE}
These advances have led to the successful application of GNNs in the design space exploration of approximate circuits, as demonstrated by Zhang et al.~\cite{Zhang2024ApproxPilot:Framework} in their ApproxPilot framework. They used manually selected features of approximate components (such as error metrics like MAE, MRE) and implemented a two-stage prediction model: (1) Node-level Classification, which predicts critical paths and integrates this information into node features, and (2) Graph-Level Regression, which uses these enhanced features for PPA (Performance, Power, Area) and accuracy prediction.

\section{Proposed Methodology} \label{sec:proposed}
In this paper, we present a construction methodology for a pre-trained graph neural network model predicting QoR and HW cost of approximate accelerators employing approximate adders from a library.

\subsection{Overview}
\begin{figure}[ht]
    \centering
    \includegraphics[width=\linewidth]{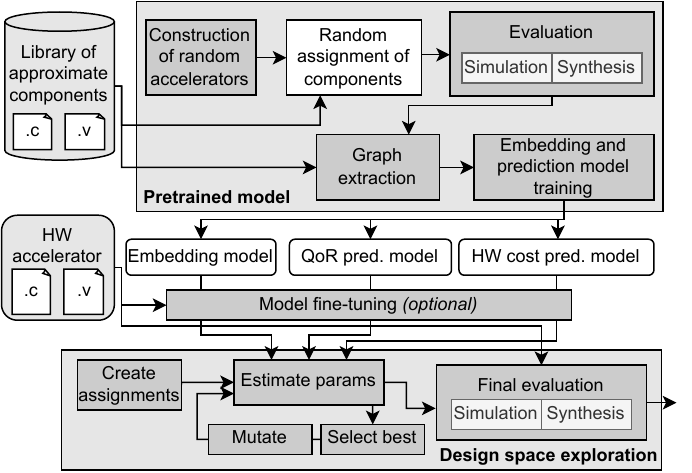}
    \caption{Overall architecture and workflow of ApproxGNN. The methodology encompasses synthetic dataset generation, graph transformation, dual-model training for component embedding and parameter prediction, and integration into design space exploration frameworks.}
    \label{fig:overall_architecture}
\end{figure}

An overview of the proposed methodology is illustrated in Fig.~\ref{fig:overall_architecture}. Our approach targets accelerators of image processing kernels constructed from the EvoApproxLib library of approximate components\cite{Mrazek2017EvoApprox8b:Methods}, providing a comprehensive workflow from circuit representation to design space exploration.

The methodology begins with \textbf{synthetic dataset generation} at the Verilog RTL level, encompassing a wide range of circuit configurations and approximate component combinations. This diverse dataset serves as the foundation for our model training process.
These Verilog descriptions are then transformed into an internal \textbf{graph representation} where nodes represent components and edges represent signal connections. This graph-based representation preserves the structural information critical for accurate parameter prediction. The graph-represented circuits are fully annotated with ground truth values for QoR and HW costs through rigorous synthesis and evaluation tools.

With the annotated graph dataset in place, we proceed to \textbf{dual-model training}. We train two complementary neural network models: a \textit{component embedding model} that generates feature representations from approximate components, capturing their functional characteristics more comprehensively than traditional error metrics \cite{Mrazek2019AutoAx:Components,Zhang2024ApproxPilot:Framework}; and a \textit{parameter prediction model} that leverages these embeddings along with circuit structure to predict QoR (approximation error) and HW costs, respectively. 

Once trained, the component embeddings can be precomputed and stored, enabling their immediate application to any accelerator incorporating those components. The prediction model can be used directly without modification or, for applications requiring higher precision, can be fine-tuned with a minimal application-specific dataset --- significantly reducing the computational overhead compared to training from scratch.

These models serve as efficient surrogate evaluators, guiding \textbf{multi-objective DSE} (NSGA-II algorithm \cite{deb2002fast}) to identify pseudo-Pareto-optimal assignments of approximate components to a fixed HW accelerator as defined in \cite{Mrazek2019AutoAx:Components,Zhang2024ApproxPilot:Framework}. The final evaluation remains necessary, as the surrogate model predictions may contain some inaccuracies at the Pareto frontier.

\subsection{Dataset generation}
Our training dataset comprises generated Verilog descriptions of approximate accelerators that implement convolutional image filter operations for various kernels (i.e., blur operations). We selected this set of tasks as representative of an error-resilient image processing application, typically solved in literature \cite{Zhang2024ApproxPilot:Framework,Mrazek2019AutoAx:Components}. The dataset generation follows a structured multi-stage process. Initially, we generate a random blur kernel with constant coefficients using a uniform random distribution ranging from 0 to a randomly selected maximum value for each kernel.

\begin{figure}[ht]
    \centering
    \includegraphics[width=0.9\linewidth]{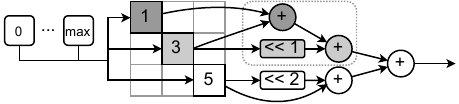}
    \caption{HW accelerator generation process example: 1) the kernel is randomly generated and 2) the shift-adder accelerator is constructed. The constant coefficient 5 is implemented as a sum of powers of 2 (1 and 4), while elements 1 and 3 are grouped.}
    \label{fig:operator_gen}
\end{figure}

Subsequently, we transform each kernel into a hardware accelerator represented as a shift-adder graph, employing a multiplierless implementation approach. Since the kernel coefficient is constant, we convert each multiplication into a series of additions and bit-shift operations, which are more hardware-efficient. To enhance dataset diversity and intentionally include sub-optimal implementations, we employ a random grouping strategy during this transformation rather than always selecting the most efficient implementation. Fig.~\ref{fig:operator_gen} illustrates this process, showing how different coefficient values (such as 5) are converted into addition operations and how elements (such as 1 and 3) can be grouped and processed together.

\begin{figure}[h!]
    \centering
    \includegraphics[width=0.95\linewidth]{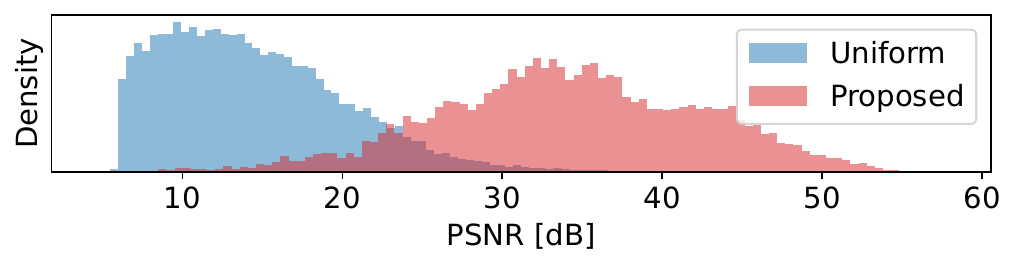}\vspace{-0.8em}%
    \caption{Distribution of PSNRs of approximate assignments for a one accelerator generated by a simple uniform sampling of the library and by the proposed weighted selection process.}\label{fig:hist}
\end{figure}

After constructing the adder graph, we perform a bit-width minimization analysis to determine the optimal bit width for each adder in the exact accelerator. In the final stage, we create hundreds of approximate accelerator variants by randomly assigning approximate adders with a desired bit-width from the EvoApproxLib library~\cite{Mrazek2017EvoApprox8b:Methods} to the generated accelerator structure. The accuracy of the accelerator is calculated as PSNR compared against the accurate implementation of the same kernel (accelerator). To ensure a more balanced approximated variant of the accelerator, we apply empirically determined weighting factors to the approximate adder component selection process preferring more accurate adders for larger accelerators, lowering bias toward very approximate implementations as shown in Fig.~\ref{fig:hist}. Uniform selection process used in \cite{Mrazek2019AutoAx:Components} generates circuits with average approximation error of 12 dB and the most accurate implementation of 32 dB, but the proposed method generates circuits with average error of 35 dB and the most accurate implementation of 55 dB.

\subsection{ML models}
\begin{figure}[ht]
    \centering
    \includegraphics[width=0.75\linewidth]{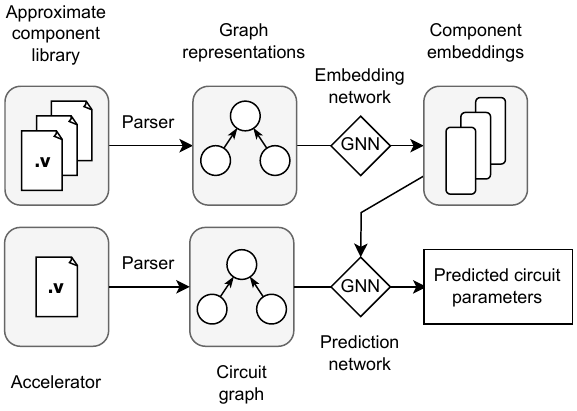}\vspace{-0.8em}
    \caption{Diagram of the proposed prediction workflow showing how component embeddings are generated and integrated with circuit representations to predict QoR and HW costs.}
    \label{fig:accuracy_workflow}
\end{figure}

Our methodology comprises three specialized neural network models: a component embedding model and two prediction models for QoR and HW cost estimation, respectively. The embedding model processes Verilog descriptions of approximate components at the gate level and generates characteristic vectors that encapsulate their functional behavior. The prediction models analyze the Verilog description of hardware accelerators in conjunction with the embeddings of their constituent approximate components to produce accurate estimates of QoR and HW costs. The relationship between these models and their integration into the overall workflow is illustrated in Fig.~\ref{fig:accuracy_workflow}.

\subsubsection{Verilog parser}
\begin{figure}[ht]
    \centering
    \includegraphics[width=0.75\linewidth]{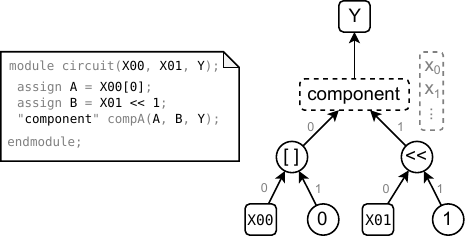}
    \caption{Example of a directed circuit graph constructed from a Verilog file with an unknown component.}
    \label{fig:graph_parse}
\end{figure}

For effective GNN-based processing, both approximate components and hardware accelerators must be transformed into graph representations. While accelerators could theoretically be generated directly in the desired graph format, approximate components are typically available only as Verilog or C++ code, necessitating a conversion process as depicted in Fig.~\ref{fig:graph_parse}.

We developed a specialized parser that converts Verilog descriptions into directed acyclic graph structures. The parser transforms hardware description language constructs such as wires, modules, and operators into nodes within a directed graph. Each node is attributed with a type that describes its specific function (e.g., different operators, constants, inputs). Our parser implements a distinctive approach to component handling: when encountering a known component, its internal structure is directly expanded and incorporated into the graph; unknown components are treated as external entities and represented using a single placeholder node --- a significant departure from approaches like \cite{Yu2021HW2VEC:Security}. These placeholder nodes contain reserved vector spaces for component features that will be populated later by our embedding model. To preserve operational semantics, edges in the graph are indexed to maintain the proper order of operands.

\subsubsection{Network architecture}

\begin{figure}[ht]
    \centering
    \includegraphics[width=\linewidth]{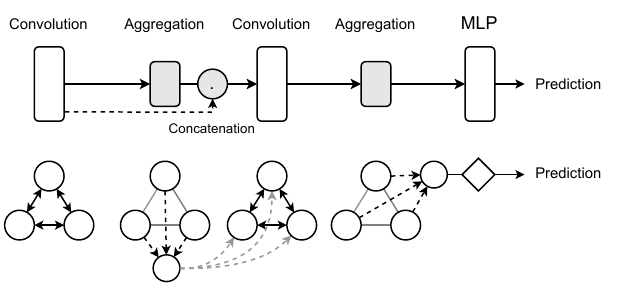}
    \caption{The proposed network architecture design illustrating the neural network layers \textbf{(top)} and their corresponding graph message exchange mechanisms \textbf{(bottom)}.}
    \label{fig:network-design}
\end{figure}

The embedding and prediction networks share a common architectural foundation, as illustrated in Fig.~\ref{fig:network-design}. Local information exchange occurs through graph convolutional layers, enabling each node to incorporate features from its neighboring nodes. This convolution operation is followed by an aggregation mechanism that collects information from the entire graph and distills it into a single representative vector. This aggregated vector is then concatenated with previous node features. The process of convolution followed by aggregation is repeated, enabling multi-hop information propagation across the graph. The second aggregated vector is processed through a multi-layer perceptron (MLP) head to generate the final prediction.

For comparative prediction tasks, such as relative accuracy between two circuits, we modify this architecture to process both inputs through identical graph processing layers except the MLP layer. This parallel processing results in two separate feature vectors that are subsequently concatenated and processed by a shared MLP head.
To enhance the network's ability to focus on critical circuit components, we employ \textit{attentional aggregation mechanisms}~\cite{li2019graphmatchingnetworkslearning} that dynamically weigh the importance of different nodes. 

\subsubsection{Model training}
A significant challenge in our approach stems from the limited diversity of the approximate component library, which typically contains Pareto-optimal implementations and is therefore restricted in size. This limitation would complicate the independent training of the embedding model due to insufficient training examples.

To address this constraint, we implement a joint training methodology where both the embedding and prediction models are trained simultaneously. During each training batch, the entire component library is processed through the embedding model to generate feature vectors. These dynamically generated embeddings are then immediately integrated as node features into the circuit graphs used for training the prediction models. This integrated approach enables backpropagation signals to flow through both models.

\section{Results}
\subsection{Experimental Setup}

We constructed three comprehensive synthetic datasets. The primary training set consists of 240 randomly generated 3x3 image convolution accelerators, with kernel coefficients selected from a uniform distribution with maximum values ranging from 5 to 20. Each accelerator was systematically transformed into a computational graph having from 8 to 24 adders and paired with 100 distinct approximate component assignments. After preprocessing and filtering, this training dataset comprises 15,900 unique circuit configurations derived from 159 different kernel patterns (hereafter referred to as the \textit{16K dataset}). The dataset is split 1:4 for validation and training. The validation dataset contains completely different accelerator architectures from the test dataset.

To assess the transfer learning capabilities of our approach, we created two additional test datasets based on standardized Gaussian filters, illustrated in Fig.~\ref{fig:used_kernels}. These filters, designated as \textit{Small Gaussian (SG)} and Large \textit{Gaussian (LG)}, were specifically selected to evaluate performance at different complexity levels. 
To ensure robust evaluation of transfer reliability, we generated 50 distinct accelerator implementations for each Gaussian filter, each with 200 component assignments. After filtering for quality and diversity, these implementations yielded 10,000 and 4,000 instances for the SG and LG datasets, respectively. We verified that the 16K training dataset contains no instances of either the SG or LG kernel patterns, ensuring a genuine test of generalization capability. All circuits of the datasets were synthesized using Synopsys Design Compiler and evaluated on 10 images to obtain PSNR between the approximated and exact implementation of the same convolutional kernel as a QoR metric.

\begin{figure}[ht]
    \centering
    \begin{subfigure}{0.5\linewidth}  
        \[\frac{1}{16} 
            \begin{bmatrix}
                1 & 2 & 1 \\
                2 & 4 & 2 \\
                1 & 2 & 1
            \end{bmatrix}
        \]
    \end{subfigure}%
    \begin{subfigure}{0.5\linewidth}  
        \[\frac{1}{255}
            \begin{bmatrix}
                16 & 31 & 16 \\
                31 & 67 & 31 \\
                16 & 31 & 16
            \end{bmatrix}
        \]
    \end{subfigure}\vspace{-0.5em}
    \caption{Small \textbf{(left)} and Large \textbf{(right)} gaussian filters used to test performance transfer.}
    \label{fig:used_kernels}
\end{figure}

We selected mean square error ($\frac{1}{|X|} \sum_{\forall_{x \in X}} (\tp{x} - \ep{x})^2$) as the main observed objective of the model accuracy. It calculates the average square difference between the estimated and true value of PSNR and power, respectively.

All experiments were conducted on a workstation with an AMD Ryzen 5 5000X CPU with 32 GB of memory. Model training and evaluation were implemented using PyTorch Geometric framework, with consistent hyperparameter settings maintained across all comparative experiments.

\subsection{QoR Model Training}
We identified QoR estimation as the most challenging aspect of our model development. We trained the prediction and embedding models according to the methodology detailed in Section \ref{sec:proposed}, implementing two distinct network architectures with parameter sizes of approximately 80 kB and 340 kB, respectively. These architectures differ in the last MLP part --- 32-64-32-1 for the small and 32-256-32-1 for the large. To establish statistical validity, we conducted 32 independent training runs for each configuration, with each complete model training for 150 epochs requiring approximately 8 minutes of computational time.

To provide a comprehensive and fair comparative analysis, we implemented two additional baseline models that represent current state-of-the-art approaches. The first baseline model, which we refer to as the \textit{metrics-based model}, follows the methodology described in \cite{Mrazek2019AutoAx:Components}, utilizing predefined error metrics of approximate components (such as mean arithmetic error, relative error, etc.) rather than learned embeddings extended by GNN instead of Random Forrest. The second baseline model, inspired by \cite{Zhang2024ApproxPilot:Framework}, focuses specifically on \textit{critical path} analysis and similarly does not employ our proposed embedding technique.

\begin{figure}[bht]
    \centering\vspace{-1em}
    \includegraphics[width=0.95\linewidth]{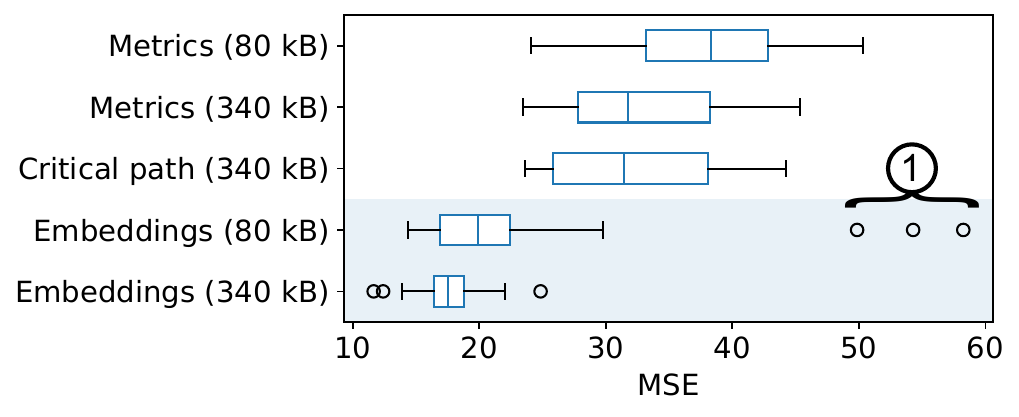}\vspace{-1em}
    \caption{Statistical comparison of validation results across 32 independent training runs per architecture including three baseline approaches.}
    \label{fig:training-runs}
\end{figure}

Fig.~\ref{fig:training-runs} presents the validation performance comparison between the error metrics-based and our embedding-based models across all 32 training runs on the 16K dataset. The results demonstrate that models leveraging our learned embeddings achieve consistently superior performance across the diverse accelerators in the dataset. 
Nevertheless, we observed that the quality of the generated embeddings with the smaller model is affected by three outliers \rpoint{1}, highlighting the importance of proper initialization. However, this variability does not occur for the larger model.


\begin{figure}[ht]
    \centering%
    \includegraphics[width=0.9\linewidth]{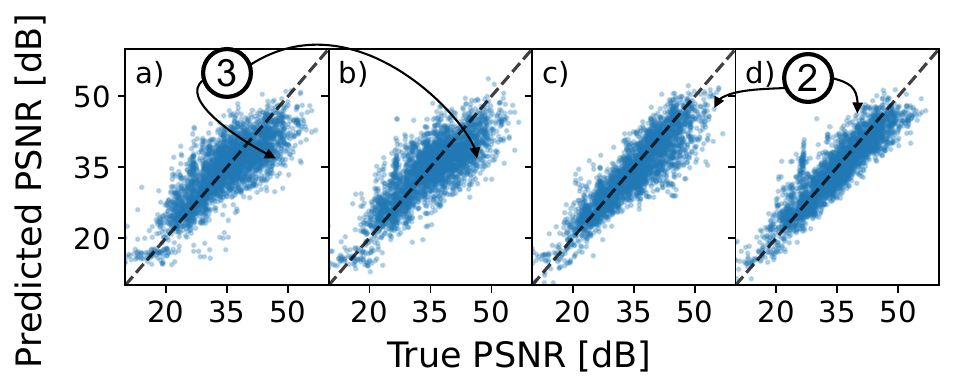}\vspace{-0.3em}
    \caption{Correlation of predicted and true values of (\textbf{a}) Metrics (340 kB), (\textbf{b}) Critical path (340 kB), (\textbf{c}) proposed embeddings (80 kB) and (\textbf{d}) proposed embeddings (340 kB) networks on 16K validation dataset.} 
    \label{fig:16k_abs_comparison}\vspace{-0.5em}
\end{figure}

Figure~\ref{fig:16k_abs_comparison} provides a more detailed analysis of the best-performing models from each category. The embedding-based model produces a more tightly clustered and symmetrically distributed set of predictions around the ideal diagonal line \rpoint{2}, indicating both higher accuracy and more balanced error distribution. In contrast, the predictions from models based on error metrics and critical path analysis exhibit a tendency to collapse toward the dataset mean \rpoint{3}, resulting in a skewed distribution.

\subsection{Transfer performance and adjustment}

\begin{figure}[b]
    \centering%
    \includegraphics[width=0.9\linewidth]{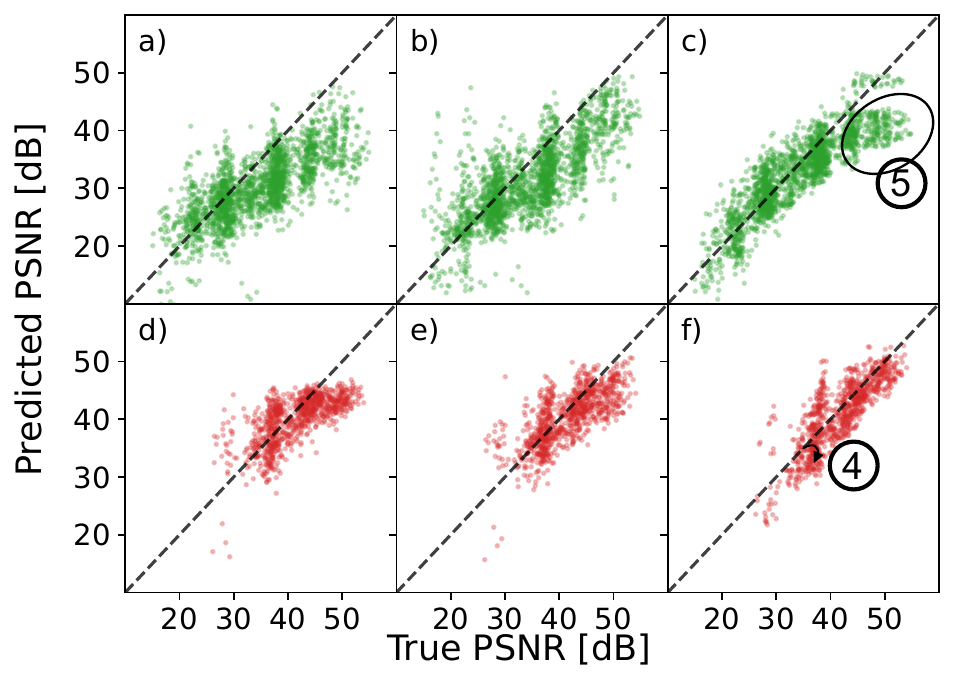}\vspace{-0.5em}%
    \caption{Correlation of predicted and true PSNR values when 16K-trained models are applied to unseen circuit configurations: SG dataset (\textbf{top}) and LG dataset (\textbf{bottom}) for error-metric model (\textbf{left}), critical-path model (\textbf{middle}) and proposed embedding-based model (\textbf{right}).}
    \label{fig:transfer_gauss}
\end{figure}

To evaluate the transferability of our approach, we applied the models previously trained on the 16K dataset to both the SG and LG datasets. For these experiments, we selected the larger 340 kB model architecture based on its performance in the aforementioned validation tests. The transfer performance results, presented in Fig.~\ref{fig:transfer_gauss}, demonstrate that the embedding-based model maintains a significantly better correlation between true and predicted PSNR compared to the metric-based and critical-path baseline models model across both test datasets. Quantitatively, the Pearson correlation coefficient improves from 0.73 and 0.75 to 0.88 for the SG dataset and from 0.67 and 0.72 to 0.83 for the LG dataset when using our embedding-based approach instead of the baseline models.

A detailed analysis of Fig.~\ref{fig:transfer_gauss} reveals that both transferred models exhibit a systematic underestimation of operator performance in both datasets \rpoint{4}. This bias is particularly visible for the LG dataset, where the predicted PSNR values are consistently lower than the actual values. The metrics-based model demonstrates a different behavior pattern when applied to the SG  dataset, producing a skewed distribution rather than a systematic offset \rpoint{5}. The pronounced underestimation observed with the Large Gaussian operators can be attributed to their substantially greater complexity than the circuits in the training set, presenting a more challenging generalization scenario.

\begin{figure}[ht]
    \centering%
    \includegraphics[width=0.95\linewidth]{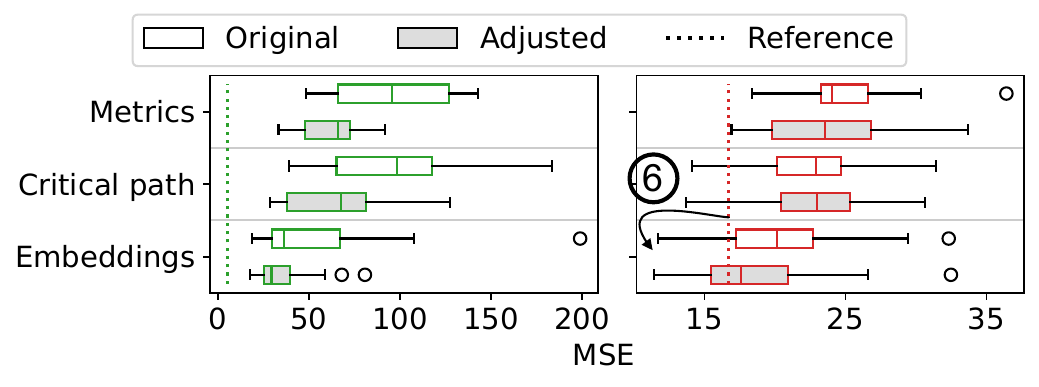}\vspace{-0.8em}%
    \caption{Quantitative comparison of transfer performance before and after applying a calibration offset derived from 10 samples. Results are shown for all 16 model instances across both the SG (\textbf{top}) and LG (\textbf{bottom}) datasets. The vertical reference lines indicate the performance of a fully trained Random Forest model on each dataset.}
    \label{fig:adjusted}
\end{figure}

To address the systematic bias in transferred models, we implemented a simple \textit{adjustment strategy} using a corrective offset. This offset was computed using 10 randomly selected samples from each target dataset. Fig.~\ref{fig:adjusted} illustrates the improvement in transfer performance across all 16 trained model instances after applying this adjustment.

For the SG dataset, both model types achieve significant improvements after adjustment, though the proposed embedding models consistently exhibit performance. The results for the LG dataset are particularly noteworthy: not only does the calibration substantially improve prediction accuracy, but all instances of our transferred and calibrated models achieve lower MSE than a Random Forest estimator that was fully trained on the LG training dataset \rpoint{6}.

\subsection{Fine-tuning and ablation study}
To identify the critical components of our QoR prediction framework, we conducted a comprehensive ablation study comparing different model architectures and training strategies. We evaluated both the critical path model (with error metric features as defined is \cite{Zhang2024ApproxPilot:Framework}) and our proposed embedding-based model, initially trained on the 16K dataset and subsequently transferred to the SG and LG datasets. For these experiments, each target dataset was partitioned into training (80\%) and validation (20\%) subsets. A 1,500 element subset of the training data was used for fine-tuning, representing 15\% of SG, 33\% of LG, and only 10\% of our original 16K dataset size --- a realistic number of assignments that is possible to evaluate. The adaptation has been done with 10 samples only.

\begin{figure}[ht]
    \centering\vspace{-0.5em}
    \includegraphics[width=0.9\linewidth]{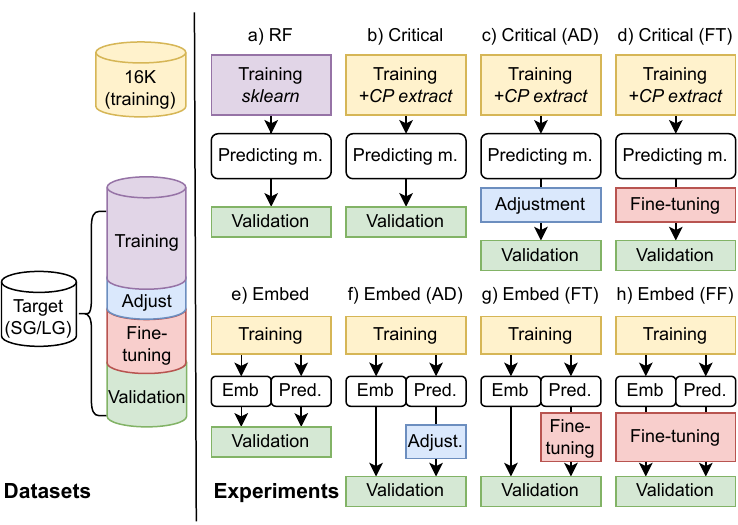}\vspace{-0.4em}
    \caption{Experimental setup for the ablation study examining different transfer and fine-tuning strategies when applying models trained on the 16K dataset to the LG and SG datasets. The color represents a model used in training, fine-tuning, adjustment, and validation steps.}
    \label{fig:experimental}
\end{figure}

Our experimental configuration, illustrated in Fig.~\ref{fig:experimental}, included several comparative scenarios: (a) a Random Forest model (RF) trained directly on the target dataset, representing the approach used in AutoAx \cite{Mrazek2019AutoAx:Components}; (b) direct application of the pre-trained Critical-path model without adaptation; (c) the transfered Critical-path model with the adjustment (AD); (d) the Critical-path model with fine-tuning (FT); (e) direct application of our Embedding-based model; (f) the Embedding model with the adjustment (AD); (g) our Embedding model with prediction network fine-tuning (FT); and (h) our Embedding model with simultaneous fine-tuning of both prediction and embedding networks (FF).

We evaluated model performance using three complementary metrics: (1) Mean Squared Error (MSE) for absolute prediction accuracy; (2) Relative MSE, computed as $\mathrm{mean}_{\forall a,b \in X}(True_\Delta(a,b) - Est_\Delta(a,b))^2$, which measures the accuracy of predicting performance differences between circuit pairs (applicable for example in \textit{distance crowding} in NSGA-II \cite{deb2002fast}); and (3) Fidelity, which quantifies how often the model correctly predicts the relative ordering ($<$, $=$, $>$) between circuit pairs, a key factor in evolutionary algorithms and Pareto front construction. We evaluated relative and fidelity metrics on a statistically significant random subset of all possible pairs ($16,000$ from $2.6\cdot10^8$), with an observed maximal variance of $\pm 0.4$ for relative MSE and $\pm 1\%$ for fidelity.

\begin{table}[b]\vspace{-0.8em}
\caption{Performance comparison across different metrics and target datasets.}\vspace{-0.5em}
\label{tab:performance}%
\newcommand{\mcx}[1]{\multicolumn{1}{c}{#1}}%
\centering
\begin{tabular}{l rr rr rr}
\toprule
& \multicolumn{2}{c}{\bf MSE} & \multicolumn{2}{c}{\bf Relative MSE} & \multicolumn{2}{c}{\bf Fidelity} \\
\cmidrule(lr){2-3} \cmidrule(lr){4-5} \cmidrule(lr){6-7}
\bf Model & \mcx{SG} & \mcx{LG} & \mcx{SG} & \mcx{LG} & \mcx{SG} & \mcx{LG} \\
\midrule
RF & 5.5 & 16.7 & 17.2 & 32.4 & 88.4~\% & 75.5~\% \\\midrule
Critical & 39.2 & 16.1 & 41.5 & 21.6 & 85.4~\% & 84.1~\% \\
Critical (AD) & 30.7 & 13.7 & 42.2 & 22.5 & 85.4~\% & 82.8~\% \\
Critical (FT) & 4.8 & 12.9 & 7.6 & 16.8 & 93.6~\% & 87.2~\% \\\midrule
Embed & 18.5 & 11.7 & 17.1 & 17.2 & 92.5~\% & 88.0~\% \\
Embed (AD) & 16.8 & 11.4 & 17.2 & 17.4 & 91.3~\% & 86.5~\% \\
Embed (FT) & 3.7 & \bf 7.6 & 4.6 & 7.9 & 93.7~\% & 91.0~\% \\
Embed (FF) & \bf 2.9 & 8.4 & \bf 3.4 & \bf 7.3 & \bf 94.8~\% & \bf 92.9~\% \\
\bottomrule
\end{tabular}
\end{table}

Table \ref{tab:performance} presents the comparative results across all model configurations. The RF model \cite{Mrazek2019AutoAx:Components} performs adequately for the simpler SG dataset but requires substantial application-specific training data. Notably, for the more complex LG dataset, our embedding-based model achieves superior results even without any retraining. Fine-tuning further improves performance for both model types, with our embedding-based approach consistently outperforming the Critical-path model proposed in \cite{Zhang2024ApproxPilot:Framework} across all metrics. The dual fine-tuning approach (FF), which updates both prediction and embedding networks, achieves the best overall performance for the SG dataset, but the transferred Embedding model (FT) is better for the more complex LG dataset. The adjustment (AD) impacts the absolute metric (MSE) only, but the pair-wise (relative and fidelity) are not affected.

These results demonstrate the efficacy of our embedding-based approach and highlight the value of transfer learning for \textbf{different image convolutional filters} in the approximate computing domain. The significant performance improvements achieved with minimal fine-tuning (using only 10\% of the data required by full training) validate our hypothesis that learned component embeddings provide a more transferable representation than traditional error metrics. Furthermore, the ablation results confirm that while fine-tuning the prediction network yields substantial improvements, jointly fine-tuning both networks provides the optimal transfer learning strategy.

\subsection{Design Space Exploration}
\begin{figure}[ht]
    \centering\vspace{-1em}
    \includegraphics[width=0.85\linewidth]{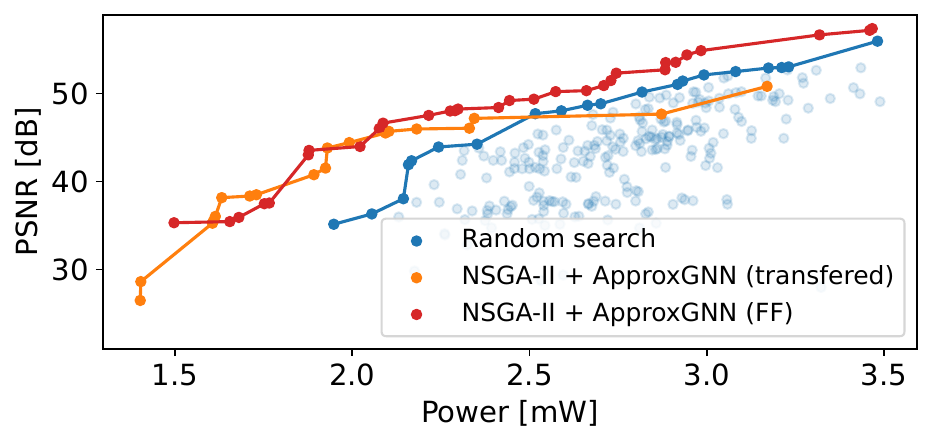}\vspace{-0.8em}
    \caption{True-power and True-PSNR of the approximate DSE of one large-gaussian filter. Comparison of NSGA-II with the proposed transferred 16k model with embedding, fine-tuned embedded model, and the random search.}\vspace{-1em}
    \label{fig:dse}
\end{figure}

In the final experiment, we deployed our 16k models as QoR and HW cost (power) predictors in a multi-objective design space exploration, assigning approximate components with surrogate models as the objectives for the best Large Gaussian filter accelerator. The transferred models exhibit PCCs 0.94 and 0.83 for power and PSNR, respectively. Additionally, the models were fine-tuned to 1000 random application-specific assignments, when PCCs improve to 0.96 and 0.98. The results presented in Fig.~\ref{fig:dse} validate that the proposed surrogate models predicted Pareto-optimal designs closely matched the true performance characteristics, mainly after the fine-tuning. The transferred models work best in the area of the middle-level approximation (below 45 dB).

\section{Conclusions}
This paper presented ApproxGNN --- a pretrained GNN for QoR and HW cost parameter of the approximated implementations of accelerator prediction. Our key contributions include a novel component embedding technique that outperforms traditional error metrics, a specialized Verilog parser for graph representation, a joint training methodology for limited component libraries, and an effective transfer learning approach. Experimental results demonstrate up to 54\% improvement in prediction accuracy compared to conventional methods, with our models achieving superior performance using just 10\% of the training data required by training from scratch \cite{Mrazek2019AutoAx:Components}. These findings validate that learned component embeddings provide more generalizable representations of approximate circuit behavior, enabling more efficient DSE for energy-efficient approximate hardware accelerators of image filters while eliminating the need for expensive retraining on each new input accelerator. This approach could also be extended to different kinds of approximate accelerators and approximate components.

\bibliographystyle{IEEEtran}
\bibliography{references}

\begin{thebibliography}{10}
\providecommand{\url}[1]{#1}
\csname url@samestyle\endcsname
\providecommand{\newblock}{\relax}
\providecommand{\bibinfo}[2]{#2}
\providecommand{\BIBentrySTDinterwordspacing}{\spaceskip=0pt\relax}
\providecommand{\BIBentryALTinterwordstretchfactor}{4}
\providecommand{\BIBentryALTinterwordspacing}{\spaceskip=\fontdimen2\font plus
\BIBentryALTinterwordstretchfactor\fontdimen3\font minus \fontdimen4\font\relax}
\providecommand{\BIBforeignlanguage}[2]{{%
\expandafter\ifx\csname l@#1\endcsname\relax
\typeout{** WARNING: IEEEtran.bst: No hyphenation pattern has been}%
\typeout{** loaded for the language `#1'. Using the pattern for}%
\typeout{** the default language instead.}%
\else
\language=\csname l@#1\endcsname
\fi
#2}}
\providecommand{\BIBdecl}{\relax}
\BIBdecl

\bibitem{axcomp}
V.~Leon, M.~A. Hanif, G.~Armeniakos, X.~Jiao, M.~Shafique, K.~Pekmestzi, and D.~Soudris, ``Approximate computing survey, part i: Terminology and software \& hardware approximation techniques,'' \emph{ACM Comput. Surv.}, vol.~57, no.~7, Mar. 2025.

\bibitem{Hanif2017QuAd}
M.~A. Hanif, R.~Hafiz, O.~Hasan, and M.~Shafique, ``{QuAd}: Design and analysis of quality-area optimal low-latency approximate adders,'' in \emph{Proceedings of the 54th Annual Design Automation Conference 2017}.\hskip 1em plus 0.5em minus 0.4em\relax New York, NY, USA: ACM, 6 2017, pp. 1--6.

\bibitem{Mrazek2017EvoApprox8b:Methods}
V.~Mrazek, R.~Hrbacek, Z.~Vasicek, and L.~Sekanina, ``{EvoApprox8b: Library of approximate adders and multipliers for circuit design and benchmarking of approximation methods},'' in \emph{Proceedings of the 2017 Design, Automation and Test in Europe, DATE 2017}, 2017.

\bibitem{Mrazek2019AutoAx:Components}
V.~Mrazek, M.~A. Hanif, Z.~Vasicek, L.~Sekanina, and M.~Shafique, ``Autoax: An automatic design space exploration and circuit building methodology utilizing libraries of approximate components,'' in \emph{Proceedings of the 56th Annual Design Automation Conference 2019}.\hskip 1em plus 0.5em minus 0.4em\relax ACM, 6 2019, pp. 1--6.

\bibitem{Zhang2024ApproxPilot:Framework}
\BIBentryALTinterwordspacing
Q.~Zhang, C.~Liu, S.~Liu, Y.~Hui, H.~Li, and X.~Li, ``{ApproxPilot: A GNN-based Accelerator Approximation Framework},'' 7 2024. [Online]. Available: \url{http://arxiv.org/abs/2407.11324}
\BIBentrySTDinterwordspacing

\bibitem{Prabakaran2023Xel-FPGAs:Systems}
B.~S. Prabakaran, V.~Mrazek, Z.~Vasicek, L.~Sekanina, and M.~Shafique, ``Xel-fpgas: An end-to-end automated exploration framework for approximate accelerators in fpga-based systems,'' in \emph{2023 IEEE/ACM International Conference on Computer Aided Design (ICCAD)}.\hskip 1em plus 0.5em minus 0.4em\relax IEEE, 10 2023, pp. 1--9.

\bibitem{Alcorta2023SpecialDesign}
E.~S. Alcorta, A.~Gerstlauer, C.~Deng, Q.~Sun, Z.~Zhang, C.~Xu, L.~W. Wills, D.~S. Lopera, W.~Ecker, S.~Garg, and J.~Hu, ``{Special Session: Machine Learning for Embedded System Design},'' in \emph{Proceedings - 2023 International Conference on Hardware/Software Codesign and System Synthesis, CODES+ISSS 2023}.\hskip 1em plus 0.5em minus 0.4em\relax Institute of Electrical and Electronics Engineers Inc., 2023, pp. 28--37.

\bibitem{Wu2022High-LevelAdvancing}
N.~Wu, H.~Yang, Y.~Xie, P.~Li, and C.~Hao, ``High-level synthesis performance prediction using gnns: benchmarking, modeling, and advancing,'' in \emph{Proceedings of the 59th ACM/IEEE Design Automation Conference}, ser. DAC '22.\hskip 1em plus 0.5em minus 0.4em\relax New York, NY, USA: Association for Computing Machinery, 2022, p. 49–54.

\bibitem{Ustun2020AccurateNetworks}
E.~Ustun, C.~Deng, D.~Pal, Z.~Li, and Z.~Zhang, ``{Accurate Operation Delay Prediction for FPGA HLS Using Graph Neural Networks},'' in \emph{IEEE/ACM International Conference on Computer-Aided Design, Digest of Technical Papers, ICCAD}, vol. 2020-November.\hskip 1em plus 0.5em minus 0.4em\relax Institute of Electrical and Electronics Engineers Inc., 11 2020.

\bibitem{Yu2021HW2VEC:Security}
S.~Y. Yu, R.~Yasaei, Q.~Zhou, T.~Nguyen, and M.~A. Al~Faruque, ``{HW2VEC: A Graph Learning Tool for Automating Hardware Security},'' in \emph{Proceedings of the 2021 IEEE International Symposium on Hardware Oriented Security and Trust, HOST 2021}.\hskip 1em plus 0.5em minus 0.4em\relax Institute of Electrical and Electronics Engineers Inc., 2021, pp. 13--23.

\bibitem{Ansari2019ImprovingMultipliers}
\BIBentryALTinterwordspacing
M.~S. Ansari, V.~Mrazek, B.~F. Cockburn, L.~Sekanina, Z.~Vasicek, and J.~Han, ``{Improving the Accuracy and Hardware Efficiency of Neural Networks Using Approximate Multipliers},'' \emph{IEEE Transactions on Very Large Scale Integration (VLSI) Systems}, 2019. [Online]. Available: \url{https://doi.org/10.1109/TVLSI.2019.2940943}
\BIBentrySTDinterwordspacing

\bibitem{openlsdfg}
L.~Ni, R.~Wang, M.~Liu, X.~Meng, X.~Lin, J.~Liu, G.~Luo, Z.~Chu, W.~Qian, X.~Yang, B.~Xie, X.~Li, and H.~Li, ``Openls-dgf: An adaptive open-source dataset generation framework for machine learning tasks in logic synthesis,'' \emph{IEEE Transactions on Computer-Aided Design of Integrated Circuits and Systems}, pp. 1--1, 2025.

\bibitem{chowdhury2021openabc}
A.~B. Chowdhury, B.~Tan, R.~Karri, and S.~Garg, ``Openabc-d: A large-scale dataset for machine learning guided integrated circuit synthesis,'' \emph{arXiv preprint arXiv:2110.11292}, 2021.

\bibitem{Li2023InvitedDesigns}
Y.~Li, M.~Liu, A.~Mishchenko, and C.~Yu, ``{Invited Paper: Verilog-to-PyG - A Framework for Graph Learning and Augmentation on RTL Designs},'' in \emph{IEEE/ACM International Conference on Computer-Aided Design, Digest of Technical Papers, ICCAD}.\hskip 1em plus 0.5em minus 0.4em\relax Institute of Electrical and Electronics Engineers Inc., 2023.

\bibitem{Wu2022dnn}
\BIBentryALTinterwordspacing
L.~Wu, P.~Cui, J.~Pei, L.~Zhao, and L.~Song, \emph{Graph Neural Networks}.\hskip 1em plus 0.5em minus 0.4em\relax Singapore: Springer Nature Singapore, 2022, pp. 27--37. [Online]. Available: \url{https://doi.org/10.1007/978-981-16-6054-2_3}
\BIBentrySTDinterwordspacing

\bibitem{Xu2022SNSsPredictor}
C.~Xu, C.~Kjellqvist, and L.~W. Wills, ``{SNS's not a Synthesizer: A Deep-Learning-Based Synthesis Predictor},'' in \emph{Proceedings - International Symposium on Computer Architecture}.\hskip 1em plus 0.5em minus 0.4em\relax Institute of Electrical and Electronics Engineers Inc., 6 2022, pp. 847--859.

\bibitem{Vaeztourshizi2023EfficientSystems}
M.~Vaeztourshizi and M.~Pedram, ``{Efficient Error Estimation for High-Level Design Space Exploration of Approximate Computing Systems},'' \emph{IEEE Transactions on Very Large Scale Integration (VLSI) Systems}, vol.~31, no.~7, pp. 917--930, 7 2023.

\bibitem{Mo2024LearningApplications}
H.~Mo, Y.~Wu, H.~Jiang, Z.~Ma, F.~Lombardi, J.~Han, and L.~Liu, ``{Learning the Error Features of Approximate Multipliers for Neural Network Applications},'' \emph{IEEE Transactions on Computers}, vol.~73, no.~3, pp. 842--856, 3 2024.

\bibitem{deb2002fast}
K.~Deb, A.~Pratap, S.~Agarwal, and T.~Meyarivan, ``A fast and elitist multiobjective genetic algorithm: {NSGA-II},'' \emph{IEEE Transactions on Evolutionary Computation}, vol.~6, no.~2, pp. 182--197, 2002.

\bibitem{li2019graphmatchingnetworkslearning}
Y.~Li, C.~Gu, T.~Dullien, O.~Vinyals, and P.~Kohli, ``Graph matching networks for learning the similarity of graph structured objects,'' in \emph{International conference on machine learning, ICML 2019}.\hskip 1em plus 0.5em minus 0.4em\relax PMLR, 2019, pp. 3835--3845.

\end{thebibliography}

\end{document}